

\def\singlespace{\normalbaselines}
\def\oneandahalfspace{\baselineskip=1.15\normalbaselineskip plus 1pt
\lineskip=2pt\lineskiplimit=1pt}

\def\np{\vfill\eject}
\def\nl{\hfil\break}

\def\nofirstpagenoten{\nopagenumbers\footline={\ifnum\pageno>1\tenrm
\hss\folio\hss\fi}}
\def\nofirstpagenotwelve{\nopagenumbers\footline={\ifnum\pageno>1\twelverm
\hss\folio\hss\fi}}
\def\leaderfill{\leaders\hbox to 1em{\hss.\hss}\hfill}
\def\ft#1#2{{\textstyle{{#1}\over{#2}}}}
\def\frac#1/#2{\leavevmode\kern.1em
\raise.5ex\hbox{\the\scriptfont0 #1}\kern-.1em/\kern-.15em
\lower.25ex\hbox{\the\scriptfont0 #2}}
\def\sfrac#1/#2{\leavevmode\kern.1em
\raise.5ex\hbox{\the\scriptscriptfont0 #1}\kern-.1em/\kern-.15em
\lower.25ex\hbox{\the\scriptscriptfont0 #2}}


\parindent=20pt
\def\narrow{\advance\leftskip by 40pt \advance\rightskip by 40pt}

\def\AB{\bigskip
        \centerline{\bf ABSTRACT}\medskip\narrow}
\def\nonarrower{\advance\leftskip by -40pt\advance\rightskip by -40pt}
\def\AE{\bigskip\nonarrower}

\def\boxit#1{\vbox{\hrule\hbox{\vrule\kern3pt
        \vbox{\kern3pt#1\kern3pt}\kern3pt\vrule}\hrule}}

\def\gtorder{\mathrel{\raise.3ex\hbox{$>$}\mkern-14mu
             \lower0.6ex\hbox{$\sim$}}}
\def\ltorder{\mathrel{\raise.3ex\hbox{$<$}|mkern-14mu
             \lower0.6ex\hbox{\sim$}}}
\def\dalemb#1#2{{\vbox{\hrule height .#2pt
        \hbox{\vrule width.#2pt height#1pt \kern#1pt
                \vrule width.#2pt}
        \hrule height.#2pt}}}

\font\fourteentt=cmtt10 scaled \magstep2
\font\fourteenbf=cmbx12 scaled \magstep1
\font\fourteenrm=cmr12 scaled \magstep1
\font\fourteeni=cmmi12 scaled \magstep1
\font\fourteenss=cmss12 scaled \magstep1
\font\fourteensy=cmsy10 scaled \magstep2
\font\fourteensl=cmsl12 scaled \magstep1
\font\fourteenex=cmex10 scaled \magstep2
\font\fourteenit=cmti12 scaled \magstep1
\font\twelvett=cmtt10 scaled \magstep1 \font\twelvebf=cmbx12
\font\twelverm=cmr12 \font\twelvei=cmmi12
\font\twelvess=cmss12 \font\twelvesy=cmsy10 scaled \magstep1
\font\twelvesl=cmsl12 \font\twelveex=cmex10 scaled \magstep1
\font\twelveit=cmti12
\font\tenss=cmss10
 
 \font\ninebf=cmbx7 scaled \magstep1
\font\ninerm=cmr7 scaled \magstep1 \font\ninei=cmmi7 scaled \magstep1
\font\ninesy=cmsy7 scaled \magstep1 
\font\eightrm=cmr7 scaled 1140 
 
\font\sevenbf=cmbx7 \font\sevenrm=cmr7 \font\seveni=cmmi7
\font\sevensy=cmsy7 

\catcode`@=11
\newskip\ttglue
\newfam\ssfam

\def\fourteenpoint{\def\rm{\fam0\fourteenrm}
\textfont0=\fourteenrm \scriptfont0=\tenrm \scriptscriptfont0=\sevenrm
\textfont1=\fourteeni \scriptfont1=\teni \scriptscriptfont1=\seveni
\textfont2=\fourteensy \scriptfont2=\tensy \scriptscriptfont2=\sevensy
\textfont3=\fourteenex \scriptfont3=\fourteenex \scriptscriptfont3=\fourteenex
\def\it{\fam\itfam\fourteenit} \textfont\itfam=\fourteenit
\def\sl{\fam\slfam\fourteensl} \textfont\slfam=\fourteensl
\def\bf{\fam\bffam\fourteenbf} \textfont\bffam=\fourteenbf
\scriptfont\bffam=\tenbf \scriptscriptfont\bffam=\sevenbf
\def\tt{\fam\ttfam\fourteentt} \textfont\ttfam=\fourteentt
\def\ss{\fam\ssfam\fourteenss} \textfont\ssfam=\fourteenss
\tt \ttglue=.5em plus .25em minus .15em
\normalbaselineskip=16pt
\abovedisplayskip=16pt plus 4pt minus 12pt
\belowdisplayskip=16pt plus 4pt minus 12pt
\abovedisplayshortskip=0pt plus 4pt
\belowdisplayshortskip=9pt plus 4pt minus 6pt
\parskip=5pt plus 1.5pt
\setbox\strutbox=\hbox{\vrule height12pt depth5pt width0pt}
\let\sc=\tenrm
\let\big=\fourteenbig \normalbaselines\rm}
\def\fourteenbig#1{{\hbox{$\left#1\vbox to12pt{}\right.\n@space$}}}

\def\twelvepoint{\def\rm{\fam0\twelverm}
\textfont0=\twelverm \scriptfont0=\ninerm \scriptscriptfont0=\sevenrm
\textfont1=\twelvei \scriptfont1=\ninei \scriptscriptfont1=\seveni
\textfont2=\twelvesy \scriptfont2=\ninesy \scriptscriptfont2=\sevensy
\textfont3=\twelveex \scriptfont3=\twelveex \scriptscriptfont3=\twelveex
\def\it{\fam\itfam\twelveit} \textfont\itfam=\twelveit
\def\sl{\fam\slfam\twelvesl} \textfont\slfam=\twelvesl
\def\bf{\fam\bffam\twelvebf} \textfont\bffam=\twelvebf
\scriptfont\bffam=\ninebf \scriptscriptfont\bffam=\sevenbf
\def\tt{\fam\ttfam\twelvett} \textfont\ttfam=\twelvett
\def\ss{\fam\ssfam\twelvess} \textfont\ssfam=\twelvess
\tt \ttglue=.5em plus .25em minus .15em
\normalbaselineskip=14pt
\abovedisplayskip=14pt plus 3pt minus 10pt
\belowdisplayskip=14pt plus 3pt minus 10pt
\abovedisplayshortskip=0pt plus 3pt
\belowdisplayshortskip=8pt plus 3pt minus 5pt
\parskip=3pt plus 1.5pt
\setbox\strutbox=\hbox{\vrule height10pt depth4pt width0pt}
\let\sc=\ninerm
\let\big=\twelvebig \normalbaselines\rm}
\def\twelvebig#1{{\hbox{$\left#1\vbox to10pt{}\right.\n@space$}}}

\def\tenpoint{\def\rm{\fam0\tenrm}
\textfont0=\tenrm \scriptfont0=\sevenrm \scriptscriptfont0=\fiverm
\textfont1=\teni \scriptfont1=\seveni \scriptscriptfont1=\fivei
\textfont2=\tensy \scriptfont2=\sevensy \scriptscriptfont2=\fivesy
\textfont3=\tenex \scriptfont3=\tenex \scriptscriptfont3=\tenex
\def\it{\fam\itfam\tenit} \textfont\itfam=\tenit
\def\sl{\fam\slfam\tensl} \textfont\slfam=\tensl
\def\bf{\fam\bffam\tenbf} \textfont\bffam=\tenbf
\scriptfont\bffam=\sevenbf \scriptscriptfont\bffam=\fivebf
\def\tt{\fam\ttfam\tentt} \textfont\ttfam=\tentt
\def\ss{\fam\ssfam\tenss} \textfont\ssfam=\tenss
\tt \ttglue=.5em plus .25em minus .15em
\normalbaselineskip=12pt
\abovedisplayskip=12pt plus 3pt minus 9pt
\belowdisplayskip=12pt plus 3pt minus 9pt
\abovedisplayshortskip=0pt plus 3pt
\belowdisplayshortskip=7pt plus 3pt minus 4pt
\parskip=0.0pt plus 1.0pt
\setbox\strutbox=\hbox{\vrule height8.5pt depth3.5pt width0pt}
\let\sc=\eightrm
\let\big=\tenbig \normalbaselines\rm}
\def\tenbig#1{{\hbox{$\left#1\vbox to8.5pt{}\right.\n@space$}}}
\let\rawfootnote=\footnote \def\footnote#1#2{{\rm\parskip=0pt\rawfootnote{#1}
{#2\hfill\vrule height 0pt depth 6pt width 0pt}}}

\def\tenfoot{\tenpoint\hskip-\parindent\hskip-.1cm}

\overfullrule=0pt
\twelvepoint
\def\sbullet{\raise.2em\hbox{$\scriptscriptstyle\bullet$}}
\nofirstpagenotwelve
\hsize=16.5 truecm
\baselineskip 15pt

\def\ft#1#2{{\textstyle{{#1}\over{#2}}}}

\def\ket#1{\big| #1\big\rangle}
\def\bra#1{\big\langle #1\big|}
\def\braket#1#2{\big\langle #1\big| #2\big\rangle}
\def\dum{{\phantom{X}}}
\def\m{{\rm -}}
\def\p{{\rm +}}

\def\phys{\big|{\rm phys}\big\rangle}

\def\deff{\Delta}
\def\teff{T^{\rm eff}}
\def\leff{L^{\rm eff}}
\def\V#1#2#3#4{{\bf V}^{#1}_{#2}[#3,#4]}
\def\aV#1#2#3#4{{\bf W}^{#1}_{#2}[#3,#4]}
\def\ie{{\it i.e.}}

\def\b{\beta}

\def\del{\partial}

\def\phys{\big|\hbox{phys}\big\rangle}

\oneandahalfspace
\rightline{CTP TAMU--86/92}
\rightline{Preprint-KUL-TF-92/43}
\rightline{hep-th/9212117}
\rightline{December 1992}

\vskip 2truecm
\centerline{\bf The Interacting $W_3$ String}
\vskip 1.5truecm
\centerline{H. Lu, C.N. Pope,\footnote{$^*$}{\tenfoot Supported in part
by the U.S. Department of Energy, under
grant DE-FG05-91ER40633.} S. Schrans\footnote{$^\diamond$}{\tenfoot
Onderzoeker I.I.K.W.;
On leave of absence from the Instituut voor Theoretische Fysica, \nl
\indent$\,$ K.U. Leuven, Belgium. Address after March 1, 1993:
Koninklijke/Shell-Laboratorium, \nl
\indent$\,$ Amsterdam (Shell Research B.V.),
Badhuisweg  3, 1031 CM Amsterdam, The Netherlands.
}
and X.J.
Wang\footnote{}{\tenfoot }}
\vskip 1.5truecm
\centerline{\it Center
for Theoretical Physics,
Texas A\&M University,}
\centerline{\it College Station, TX 77843--4242, USA.}

\vskip 1.5truecm
\AB\singlespace
   We present a procedure for computing gauge-invariant scattering
amplitudes in the $W_3$ string, and use it to calculate three-point and
four-point functions.  We show that non-vanishing scattering amplitudes
necessarily involve external physical states with excitations of ghosts as
well as matter fields.  The crossing properties of the four-point functions
are studied, and it is shown that the duality of the Virasoro string
amplitudes generalises in the $W_3$ string, with different sets of
intermediate states being exchanged in different channels.  We also exhibit
a relation between the scattering amplitudes of the $W_3$ string and the
fusion rules of the Ising model.
\AE\oneandahalfspace

\np
\noindent
{\bf 1. Introduction}
\bigskip

     $W$ strings are a natural generalisation of the ordinary bosonic string
[1--6].  The spectrum of physical states of a $W$-string theory is given by
the non-trivial cohomology of its BRST operator.  Since the BRST operator is
known explicitly only for the case of $W_3$ [7], most attention has focussed
on this example.  Considerable progress has been made in determining the
spectrum of physical states of the $W_3$ string [2,4,5,8,9,10] but until now,
very little progress has been made in introducing interactions in the
theory.  The basic reason for this is that there seems to be no way of
constructing non-vanishing gauge-invariant correlation functions involving
only physical states of standard ghost structure.  In this paper we show
how such correlation functions can in fact be built. The new ingredient that
allows this is the existence, and recent discovery [9,10], of new physical
states with non-standard ghost structure, which have excitations of the
ghosts as well as of the matter fields. Non-vanishing scattering amplitudes
necessarily require that some external physical states have this form. Thus
the $W_3$-string theory is rather unusual in that such states seem to be
essential in constructing an interacting theory.

     For the 26-dimensional bosonic string, the physical states all have the
form $\ket{\psi}=\ket{X} \otimes \ket{\rm gh}$, where $\ket{X}$ is built
exclusively from creation operators of the 26 matter fields $X^\mu$ acting on
a momentum eigenstate, and $\ket{\rm gh}$ is the standard ghost vacuum.  For
the multi-scalar $W_3$ string analogous physical states also exist, and have
been fully classified [4,5,8].  However, it has recently been appreciated
that this does not exhaust the non-trivial cohomology of the BRST operator
for the $W_3$ string [10].  In fact there are many additional physical
states, which have a non-standard structure in that they involve excitations
of the ghost as well as the matter fields.  The occurrence of these states
can be attributed to the fact that there does not exist a physical gauge for
the $W_3$ string.  For example, there are massive vector states in the $W_3$
string, indicating the existence of less spacetime gauge symmetry than for
the ordinary bosonic string.  The phenomenon of physical states with
non-standard ghost structure in fact occurs in the bosonic string, but only
in the special case when there are two spacetime dimensions [11,12].
However, physical states with non-standard ghost structure occur in the {\it
multi-scalar} $W_3$ string [10], where the on-shell momenta take {\it
continuous} values, by contrast to the two-scalar bosonic string where the
momenta are necessarily discrete.

     The purpose of this paper is to develop a procedure for computing
gauge-invariant scattering amplitudes for the $W_3$ string, and to discuss
their significance.  In section 2, we present a short review of the
multi-scalar $W_3$ string and the construction of its physical states.  The
construction is exhaustive for  physical states with the standard ghost
structure.  The spectrum of  non-standard physical states has not yet been
obtained completely, but  sufficiently many examples are now known to enable
us to investigate some of  the features of the interacting theory.  The
explicit forms of the relevant physical states are relegated to an appendix.

     In section 3, we present our procedure for writing down gauge-invariant
scattering amplitudes for the $W_3$ string.  We shall illustrate the
procedure by computing specific examples of three-point and four-point
tachyon scattering amplitudes.  In section 4 we discuss the
interpretation of these amplitudes, including their crossing and duality
properties and their relation to the Ising model. The paper ends with
conclusions and comments in section 5.

\bigskip
\bigskip
\noindent
{\bf 2. Physical states of the multi-scalar $W_3$ string}
\bigskip

     The key ingredient for determining the physical spectrum of the $W_3$
string is the construction of the BRST operator [7], which is given by
$$
Q_B=\oint dz \Big[c\,(T+\ft12 T_{\rm gh})+\gamma\,(W+\ft12
W_{\rm gh})\Big], \eqno(2.1)
$$
and is nilpotent provided that the matter currents $T$ and $W$ generate the
$W_3$ algebra with central charge $c=100$, and that the ghost currents are
chosen  to be
$$
\eqalignno{
T_{\rm gh}&=-2b\,\partial c-\partial
b\, c-3\beta\, \partial\gamma-2\partial\beta\, \gamma\ , &(2.2)\cr
W_{\rm gh}&=-\partial\beta\,
c-3\beta\, \partial c-\ft8{261}\big[\partial(b\, \gamma\,  T)+b\,
\partial\gamma \, T\big]\cr
&\ \ +\ft{25}{1566}\Big(2\gamma\, \partial^3b+9\partial\gamma\,
\partial^2b +15\partial^2\gamma\,\partial b+10\partial^3\gamma\,
b\Big) ,&(2.3)\cr}
$$
where the ghost-antighost pairs ($c$, $b$) and ($\gamma$, $\beta$) correspond
respectively to the $T$ and $W$ generators. A matter
realisation of $W_3$ with central charge 100 can be given
in terms of $n\ge 2$ scalar fields, as follows [13]:
$$
\eqalign{
T&= -\ft12 (\del\varphi)^2 - Q \del^2 \varphi +\teff,\cr
W&=-{2i \over \sqrt{261} }\Big[ \ft13 (\del\varphi)^3 + Q \del\varphi
\del^2\varphi +\ft13 Q^2 \del^3 \varphi + 2\del\varphi \teff
+ Q \del \teff\Big],\cr}\eqno(2.4)
$$
where $Q^2=\ft{49}{8}$ and $\teff$ is an energy-momentum tensor
with central charge $\ft{51}2$ that commutes with $\varphi$.  Since
$\teff$ has a fractional central charge, it cannot be realised simply by
taking free scalar fields.  We can however use $d$ scalar fields $X^\mu$
with a background-charge vector $a_\mu$:
$$
\teff=-\ft12 \del X_\mu \del X^\mu - i a_\mu \del^2 X^\mu,\eqno(2.5)
$$
with $a_\mu$ chosen so that $\ft{51}2=d - 12 a_\mu a^\mu$ [4].

     Physical states are by definition states that are annihilated by the BRST
operator (2.1) but that are not BRST trivial. We shall first consider such
states  with standard ghost structure, {\it i.e.}
$$
\ket{\chi}=\ket{\varphi\, ,X}\otimes\ket{\m\m}.\eqno(2.6)
$$
Here $\ket{\m\m}$ is the standard ghost vacuum, given by
$$
\ket{\m\m}=c_1\gamma_1\gamma_2\ket{0}.\eqno(2.7)
$$
The $SL(2,C)$ vacuum satisfies
$$
\eqalignno{
c_n\ket{0}=0,\qquad n\ge 2;&\qquad\qquad\qquad
b_n\ket{0}=0,\qquad n\ge -1,&(2.8a)\cr
\gamma_n\ket{0}=0,\qquad n\ge 3;&\qquad\qquad\qquad
\beta_n\ket{0}=0,\qquad n\ge -2.&(2.8b)\cr}
$$
The anti-ghost fields $b$,
$\b$ have ghost number $G=-1$, and the ghost fields $c$, $\gamma$ have
ghost number $G=1$.\footnote{$^*$}{\tenfoot For states, we adopt the
convention that the ghost vacuum $\ket{\m\m}$ has ghost number $G=0$, which
means that the $SL(2,C)$ vacuum $\ket{0}$ has ghost number $G=-3$. This
implies that physical states of ghost number $G$ are obtained by acting on
$\ket{0}$ with operators of ghost number $(G+3)$. }

     For standard states of the form (2.6), the condition of BRST invariance
becomes [7]:
$$
\eqalign{
(L_0-4)\ket{\varphi\, ,X}&=0,\cr
W_0\ket{\varphi\, ,X}&=0,\cr
L_n\ket{\varphi\, ,X}=W_n\ket{\varphi\, ,X}&=0,\qquad n\ge 1.\cr}\eqno(2.9)
$$
The consequences of these physical-state conditions have been studied in
detail in various papers [2,4,5,8].  The main features that emerge are the
following. The excited states can be divided into two kinds, namely those
for which there are no excitations in the $\varphi$ direction, and those
where $\varphi$ is excited too. The latter states are all null, as has been
discussed in [5,8], and thus need not be considered further.  For the
former, we may write $\ket{\varphi\, ,X}$ as
$$
\ket{\varphi\, ,X}=  e^{\beta\varphi(0)}{\phys}_{\rm eff},\eqno(2.10)
$$
where ${\phys}_{\rm eff}$ involves only the $X^\mu$ fields and not
$\varphi$.
The physical-state conditions (2.9) now imply that
$$
(\beta+Q)(\beta+\ft67 Q)(\beta+\ft87 Q)=0,\eqno(2.11)
$$
together with the effective physical-state conditions [2,4,5]:
$$
\eqalign{
(\leff_0 - \deff){\phys}_{\rm eff}&=0,\cr
\leff_n{\phys}_{\rm eff}&=0,\qquad n\ge 1.\cr}\eqno(2.12)
$$
The value of the effective intercept $\deff$ is 1 when
$\beta=-\ft67 Q$ or $-\ft87 Q$, and it equals $\ft{15}{16}$ when $\beta=-Q$.
Thus these states of the $W_3$ string are described by two effective
Virasoro-string spectra, for an effective energy-momentum tensor
$\teff$ with central charge $c=\ft{51}2$ and intercepts $\deff=1$
and $\deff=\ft{15}{16}$. The first of these gives a mass
spectrum similar to an ordinary string, with a massless vector at level 1,
whilst the second gives a spectrum of purely massive states [5].

    As we have indicated in the introduction, the $W_3$-string spectrum is
much richer than simply that of the standard physical states we have
discussed so far. Although the classification of physical states with
non-standard ghost structure is as yet incomplete, some classes of such
states have been found [10,14].  They contain excitations of the ghost and
antighost fields as well as the matter fields.  The level number $\ell$ of
these states is defined with respect to the ghost vacuum $\ket{\m\m}$ given
in (2.7).  Thus, for example, the $SL(2,C)$ vacuum $\ket{0}$ has level
number $\ell=4$ and ghost number $G=-3$, since it can be written as
$\beta_{-2} \beta_{-1} b_{-1}\ket{\m\m}$.  It is straightforward to see that
at level $\ell$, the allowed ghost numbers of {\it states} (not
necessarily physical) lie in the interval
$$
\big[ 1-\sqrt{4\ell+1}\big] \le G \le \big[ 1+\sqrt{4\ell+1}\big],
\eqno(2.13)
$$
where $\big[a\big]$ denotes the integer part of $a$. In fact for all known
examples of {\it physical} states [10], the ghost number lies in the restricted
interval $-3\le G \le 5$.

     All the physical states in the $W_3$ string occur in
multiplets [10], each of which may be viewed as being built from what we shall
call a {\it prime state}.  We shall discuss the prime states first.  All the
known examples of prime states occur either at ghost number $G=0$ (in the
case of states with standard ghost structure), or
 at $G=-1$ or
$G=-3$ (in the case of states with non-standard ghost structure).
The corresponding  operators have ghost numbers $G=3$, $G=2$ or $G=0$.
{}From (2.13), we see that the last of these can arise only when the
level number $\ell$ is $\ge 4$, and thus we first discuss the $G=-1$ prime
states.  (In fact the examples that we shall be concerned with in this paper
involve level numbers $\le 3$.)

     The first non-standard physical states arise at level $\ell=1$.  They
are given by operators of the form
$$
V=\Big( c\,\gamma \mp {i\over 3 \sqrt{58}} \del \gamma\,\gamma \Big) e^{\beta
\varphi +i p\cdot X} \eqno(2.14)
$$
acting on the $SL(2,C)$ vacuum $\ket{0}$ [10].  Just as in the case of the
physical states with standard ghost structure, here the corresponding physical
states, and indeed all the non-standard physical states we shall consider in
this paper, admit a natural interpretation as string-like states for the
effective spacetime described by the coordinates $X^\mu$. The operator $V$
in (2.14) describes a physical state if $\beta=-\ft47Q$ and the $-$ sign is
chosen, or if $\beta=-\ft37 Q$ and the $+$ sign is chosen.  The former
operator leads to a state with effective spacetime intercept
$\leff_0=\deff=\ft12$, whilst the latter corresponds to $\deff=\ft{15}{16}$.
We denote these physical operators by $\V{G}{\deff}{\beta}{p}=
\V{2}{1/2}{-\ft47 Q}{p}$ and
$\V{2}{15/16}{-\ft37 Q}{p}$, and shall adopt similar conventions for the
other physical states. The spectrum of the $G=-1$ level 1 prime states  is
completed by two discrete states, with $\beta$ momenta $-\ft67Q$ and
$-\ft87Q$, and spacetime momentum $p_\mu=0$ in each case [10].

     At level $\ell=2$, there is just one physical state with non-standard
ghost structure at $G=-1$.  It has momentum $\beta=-\ft27 Q$ in the
$\varphi$ direction, and effective spacetime intercept $\deff=\ft12$.
Following the same notation as above, we denote the corresponding $G=2$
operator by $\V{2}{1/2}{-\ft27 Q}{p}$.  Its detailed form is given in (A.11).

     At level $\ell=3$, there are three $G=-1$ physical states with
non-standard ghost structure.  There is a state with $\beta=0$ and
$\deff=1$; we denote the corresponding operator by $\V{2}{1}{0}{p}$.  Its
detailed form is given in (A.12).  Since in this paper we shall not need the
two other prime states that occur at this level, we shall not give
their explicit form, but merely  note that they have $\beta=-\ft47
Q$ and $\beta= -\ft37 Q$.  A more detailed discussion of physical states
with  non-standard ghost structure will be given in [15].

     Prime states with $G=-1$ presumably exist at all higher level numbers
too, but as yet no classification of them exists.  There are also further
prime states at $G=-3$, [10], which, in view of (2.13), can occur only when the
level number satisfies $\ell\ge 4$.  In fact the first example, at $\ell=4$,
is the $SL(2,C)$ vacuum $\ket{0}=\beta_{-2}\beta_{-1}b_{-1}\ket{\m\m}$.
This is manifestly BRST invariant and BRST non-trivial, and thus provides a
zero-momentum discrete state.

     Higher-level $G=-3$ physical states were first found in [10], in the
context of the two-scalar $W_3$ string, where all physical states have
discrete momenta.   Some physical states in the two-scalar $W_3$ string
generalise to continuous-momentum physical states in the multi-scalar
$W_3$ string, some generalise to discrete-momentum physical states, and some
do not admit generalisations at all.  We have seen examples above of states
in the first two categories.  At level $\ell=3$ there are further discrete
states in the two-scalar $W_3$ string that do not generalise to the
multi-scalar case.  In [10], two $\ell=6$ discrete states at $G=-3$ were found
for the two-scalar $W_3$ string; these are analogues of the ground-ring
generators [12] for the discrete states of the two-scalar bosonic string.  One
of these $\ell=6$ states generalises to a discrete state in the multi-scalar
$W_3$ string; it corresponds to a $G=0$ operator, with $\beta=\ft27Q$ and
spacetime momentum $p_\mu=0$.  In [16], four further discrete states in the
two-scalar $W_3$ string were found, at $\ell=8$ and $G=-3$.  Together with
the $\ell=6$ states described above, they constitute the complete set of
``ring generators'' for the two-scalar $W_3$ string [16]. Two of them (a
pair with conjugate values of momentum in the second direction) generalise to
give a continuous-momentum physical state in the multi-scalar $W_3$ string,
corresponding to a $G=0$ operator with $\beta=\ft47 Q$ and continuous
on-shell spacetime momentum. The detailed expressions for the $\ell=6$ and
$\ell=8$ operators that we have described here are quite complicated.  Since
we shall not be making use of them in the present paper, we shall not give
their explicit form. Details may be found in [10,16].

     As we mentioned above, the physical states with non-standard ghost
structure arise in multiplets [10], and so far we have described just the {\it
prime states}.  For each multiplet, this is the state from which all the
other multiplet partners may be generated.  The multiplet is then
constructed by acting on the prime state with the $G=1$ operators $a_\varphi
\equiv [Q_B,\varphi]$ and $a_{X^\mu}^\dum \equiv [Q_B, X^\mu]$.  These
operators are manifestly BRST invariant, but they are not BRST trivial since
$\varphi$ and $X^\mu$ are not primary conformal fields [12,10].  The
$a_\varphi$
and $a_{X^\mu}^\dum$ operators, which have conformal weight 0, act by normal
ordering with the operator that creates the physical state, giving rise in
general to new physical states with ghost number boosted by 1.  By
repeatedly acting with $a_\varphi$ and $a_{X^\mu}^\dum$, the entire
multiplet associated with a given prime state is generated.  In the present
paper, the only such physical states that we shall be concerned with are
those obtained by acting just once on a prime state, \ie\ states with ghost
number $G=0$.  For convenience, we summarise the prime states that we shall
be using in this paper in a table:

\bigskip
$$
\hbox{
\vbox{\tabskip=0pt \offinterlineskip
\def\tablerule{\noalign{\hrule height1pt}}
\halign to250pt{\strut#& \vrule#\tabskip=0em plus10em&
\hfil#\hfil& \vrule#& \hfil#\hfil& \vrule#& \hfil#\hfil& \vrule#&
\hfil#& \vrule#\tabskip=0pt\cr\tablerule
&&\phantom{}&&${\scriptstyle G}$&&${\scriptstyle L_0^{\rm eff}}$
&&${\scriptstyle \beta\ \qquad}$&\cr\tablerule
&&\phantom{}&&${\scriptstyle 3}\,$&&${\scriptstyle 15/16 }$
&&${\scriptstyle -Q}\ \qquad$&\cr
&&$\ell=0$
&&${\scriptstyle 3}$
&&${\scriptstyle 1}$
&&${\scriptstyle -6Q/7, \quad -8Q/7}$&\cr
\tablerule
&&\phantom{}&&${\scriptstyle 2}\,$&&${\scriptstyle 15/16 }$
&&${\scriptstyle -3Q/7}$\qquad&\cr
&&$\ell=1$
&&${\scriptstyle 2}$
&&${\scriptstyle1/2}$
&&${\scriptstyle -4Q/7}\qquad$&\cr
\tablerule
&&$\ell=2$
&&${\scriptstyle 2}\,$&&${\scriptstyle 1/2}$
&&${\scriptstyle -2Q/7}\qquad$&\cr \tablerule
&&$\ell=3$
&&${\scriptstyle 2}\,$&&${\scriptstyle 1}$&&${\scriptstyle 0}\qquad\quad$&\cr
\tablerule \noalign{\bigskip}}}}
$$
\centerline{\it Table 1.}
\bigskip

     Every BRST non-trivial physical state has non-zero norm, in the sense that
it has a non-vanishing inner product with some other physical state, \ie\
its conjugate.  We shall defer a more detailed discussion of this until the
next section. For now, we just remark that the conjugate of a state with
ghost number $G$ and momentum $(\beta,p_\mu)$ occurs at ghost number $-G+2$
and, because of the background charges, at momentum $(-\beta-2Q,
-p_\mu-2a_\mu)$.

     To conclude this section, we remark that {\it all} the BRST non-trivial
physical states are highest-weight states with conformal dimension 0 with
respect to the total energy-momentum tensor $T^{\rm tot}\equiv T+T_{\rm
gh}$. This follows because, as in ordinary string theory [12], any physical
state $\ket{\chi}$ is certainly an eigenstate of $L_0^{\rm tot}$, and so we
have $L_0^{\rm tot}\ket{\chi}=\lambda \ket{\chi}=\{Q_B,b_0\}\ket{\chi} = Q_B
b_0\ket{\chi}$.  Thus if $\ket{\chi}$ is BRST non-trivial, it must be that
$\lambda=0$, since otherwise we would have $\ket{\chi}=\lambda^{-1} Q_B
b_0\ket{\chi}$. \footnote{$^*$}{\tenfoot It is not true, however, that there
exists in general a basis for the physical states such that they are
eigenstates of $W^{\rm tot}_0$, even though $Q_B$, $L^{\rm tot}_0$ and
$W^{\rm tot}_0$ commute.  To see this, suppose that $\ket{\chi_i}$ denotes
{\it all} the states at given $\ell$ and $G$ that are annihilated by $Q_B$.
Since $Q_B W^{\rm tot}_0\ket{\chi_i}=0$, it follows that we can write
$W^{\rm tot}_0\ket{\chi_i}=\sum_j a_{ij}\ket{\chi_j}$.  The problem is that
$a_{ij}$ is non-hermitean, and cannot always be diagonalised, even though
$W^{\rm tot}_0$ is an hermitean operator.  (This is because $a_{ij}$ are not
the matrix elements of $W^{\rm tot}_0$ with respect to the proper
$SL(2,C)$-invariant inner product.) We have found explicit examples of
physical states at $\ell=4$ and $G=0$ where the diagonalisation of $a_{ij}$
is impossible.  It is true, however, that $W^{\rm tot}_0$ on any physical
state gives a BRST-trivial state, since $W^{\rm tot}_0\ket{\chi}
=\{Q_B,\beta_0\}\ket{\chi}=Q_B\beta_0\ket{\chi}$. It follows from this that
the matrix of $W^{\rm tot}_0$ inner products with respect to the proper
$SL(2,C)$-invariant inner product is not merely hermitean, but actually
zero.}

\np
\noindent
{\bf 3. Interactions in the $W_3$ string}
\bigskip

\noindent{\it 3.1 Introduction}
\bigskip

     As in ordinary string theory at the first-quantised level [17],
interactions for the $W_3$ string have to be introduced by hand. The guiding
principle for the construction of the interaction terms is that they should
be gauge invariant. In other words, we need to build BRST invariant scattering
amplitudes. These can be constructed from correlation functions of the BRST
invariant physical operators that we described in section 2. However, as we
shall explain below, there are no interactions among physical states with only
standard ghost structure. The main point of this paper is to show that
interactions can in fact occur in the $W_3$ string, but that they necessarily
involve states of non-standard ghost structure.

     For a correlation function of operators to be non-vanishing, two
necessary conditions must be satisfied. The first is that the operators must
have the correct total ghost number; the second is that they must satisfy
momentum conservation.

   In addition to the standard ghost vacuum $\ket{\m\m}$ defined in (2.7),
there are three more ghost vacua $\ket{\p\m}=c_0\ket{\m\m}$,
$\ket{\m\p}=\gamma_0\ket{\m\m}$ and $\ket{\p\p}=c_0\gamma_0\ket{\m\m}$ which
are degenerate in energy with $\ket{\m\m}$. Since we therefore have that
$\ket{\m\m}=\beta_0 b_0\ket{\p\p}$, it follows that $\braket{\m\m}{\m\m}=
\braket{\p\p}{\p\p}=0$, {\it etc.} The basic non-vanishing inner product is
$$
\eqalign{
1=\braket{\p\p}{\m\m}&=\bra{0} c_{-1}c_0c_1\,\gamma_{-2}\gamma_{-1}\gamma_0
\gamma_1\gamma_2\ket{0}\cr
&=\ft1{576}\bra{0} \del^2 c\,\del c \,c\,
\del^4\gamma\,\del^3\gamma\,\del^2\gamma\,\del\gamma\,\gamma\ket{0}\ .\cr}
\eqno(3.1)
$$
Note that in particular the total ghost number of the operators in a
non-vanishing correlator must be $3+5=8$.

     We find it convenient for calculating correlation functions to bosonise
the ghosts as follows:
$$
\eqalign{
b&=e^{-i\sigma};\quad c=e^{i\sigma}\ ,\cr
\beta&=e^{-i\rho};\quad \gamma=e^{i\rho}\ ,\cr}\eqno(3.2)
$$
where $\sigma$ and $\rho$ are real scalars with the operator-product
expansions $\sigma (z)\sigma
(w)\sim -\log (z-w)$ and $\rho (z)\rho (w)\sim -\log (z-w)$. It is
straightforward to see that
$$
\del^n c\,\del^{n-1}c\,\cdots \,\del c \,c=n!(n-1)!\cdots 1\,
e^{i(n+1)\sigma}\ , \eqno(3.3)
$$
and similarly for $\gamma$. Thus (3.1) becomes
$$
\bra{0} \,e^{3i\sigma}\, e^{5i\rho}\,\ket{0}=1\ .\eqno(3.4)
$$

     In addition to having the ghost structure given above, non-vanishing
correlators must also satisfy momentum conservation. Owing to the presence
of the background charges, we must have $\sum_{i=1}^N p^\mu_i =-2a^\mu$ in
the effective spacetime together with
$$
\sum_{i=1}^N \beta_i=-2Q \eqno(3.5)
$$
in the $\varphi$ direction, in order to have a non-vanishing $N$-point
function. For states of continuous spacetime momentum $p_\mu$, as indeed we
have in the multi-scalar $W_3$ string,  momentum conservation in the $X^\mu$
directions can be straightforwardly  satisfied. However, as we saw in
section 2, the momentum $\beta$ in the $\varphi$ direction can only take
specific frozen values in physical states.  Thus it is in general
non-trivial to satisfy momentum conservation in the  $\varphi$ direction.

     Note {\it en passant} that since all the physical operators are primary
fields with dimension zero with respect to $T+T_{\rm gh}$, the conformal
prefactors in all correlation functions will be trivial.

\bigskip
\noindent{\it 3.2 The two-point function}
\bigskip

    We begin our detailed discussion of correlation functions by considering
the two-point function. It is this that defines the inner product, and hence
the meaning of conjugation of any state. Physical states with standard ghost
structure provide a good example for this discussion. One can see from
(2.6), (2.7) and (2.10) that the $G=3$ physical operator describing a
standard  state is of the form
$$
\eqalign{
\V{3}{\deff}{\beta}{p}&=c\,\del\gamma\,\gamma\, e^{\beta\varphi}\,
P(\del X) e^{ip\cdot X}\cr
&=e^{i\sigma}\, e^{2i\rho}\,e^{\beta\varphi}\, P(\del X)e^{ip\cdot X}\ ,\cr}
\eqno(3.6)
$$
where $\beta$ satisfies (2.11), $P(\del X)e^{ip\cdot X}\ket{0}=\phys_{\rm
eff}$ satisfies (2.12), and $\Delta$ is given below (2.12).  The
$G=5$ operator conjugate to (3.6) that acts on $\bra{0}$ to create the
conjugate state has the form
$$
\eqalign{
\V{5}{\deff'}{\beta'}{p'}&=
\del c\, c\,\del^2\gamma\,\del\gamma\,\gamma\, e^{\beta'\varphi} \,
P(\del X)e^{ip'\cdot X}\cr
&=2\, e^{2i\sigma}\, e^{3i\rho}\,e^{\beta'\varphi} \,P(\del X)
e^{ip'\cdot X}\ .\cr} \eqno(3.7)
$$
The physical-state conditions imply that $\beta'$ must satisfy the same
cubic equation (2.11) as does $\beta$, and $P(\del X)e^{ip'\cdot X}$ gives
rise to a state satisfying (2.12). From (3.5) we see that the solution
$\beta=-Q$ of (2.11) is self-conjugate, and that the solutions
$\beta=-\ft67Q$ and $\beta=-\ft87Q$ are conjugate to each other [4,5,8]. The
inner
product between the above states, when we choose $\beta'$ so as to
satisfy (3.5), is
$$
\eqalign{
&\bra{0}\V{5}{\deff}{-2Q-\beta}{-2a- p}(z_1)\;
\V{3}{\deff}{\beta}{p}(z_2)\ket{0}\cr
&=2\, \bra{0} e^{2i\sigma} \,e^{3i\rho}\,e^{(-2Q-\beta)\varphi}(z_1)\;
e^{i\sigma}\, e^{2i\rho}\,e^{\beta\varphi}(z_2)\ket{0}
\braket{\rm phys'}{\rm phys}_{\rm eff}\cr
&=-2\, z_{12}^2\ z_{12}^6\ z_{12}^{\beta (\beta+2Q)}\ z_{12}^{-2\Delta}\ .\cr}
\eqno(3.8)
$$
In the last line we have used Wick's theorem and written the contributions
from $\sigma$, $\rho$, $\varphi$ and $\phys_{\rm eff}$ in that order. We
 use the standard notation $z_{ij}\equiv z_i-z_j$.  For
each solution given by (2.11) and its corresponding value of $\deff$ given
below (2.12), we see that the exponent of $z_{12}$ is zero, and thus the
inner product is just a constant, as it must be for the two-point function
of conformal-weight 0 fields.  (The precise value of the constant is
determined by the normalisations of the physical operators.  Of course
because the inner product is off-diagonal one must, as in ordinary string
theory, truncate the states to obtain a positive-definite metric on the
Hilbert space [8].  This is usually done by identifying states and their
conjugates.)  The calculation of inner products of states involving
non-standard ghost structures proceeds in a similar fashion.

\bigskip
\noindent{\it 3.3 The three-point function}
\bigskip

     In ordinary string theory the three-point function defines the basic
interaction vertex of the theory [17].  This is also the case for the $W_3$
string, as we shall explain.  There are, however, differences and subtleties
that do not arise for the ordinary string.

     Three-point functions are identically zero if all the external states
have the standard ghost structure (2.6).  This follows from the fact that
neither of the two necessary conditions discussed in subsection 3.1 is
satisfied in this case. To see this, we note that the ghost number of the
physical operator that creates a standard state is $G=3$, and thus the
product of  three such operators has ghost number $G=9$.  From (3.1) it then
follows that the inner product in the ghost sector gives zero.  Note that
since the conjugate operators have ghost number $G=5$, they cannot remedy
this problem.  There is also another reason why these three-point functions
vanish.  Since the allowed values of $\beta$ momentum are in this case
$\beta=-Q$, $\beta=-\ft67Q$ or $\beta=-\ft87Q$, it follows that three of
them cannot be combined so as to satisfy the momentum-conservation law
(3.5).  Similar arguments show that all higher $N$-point functions of
physical states with the standard ghost structure, or their conjugates, are
identically zero.

     This apparent difficulty of introducing interactions in $W_3$ string
theory can be resolved by considering three-point functions of physical
states with non-standard ghost structure.  These states circumvent both of
the  difficulties described above, since, as we saw in section 2, they occur
with  lower ghost number and less negative $\beta$ momentum, as compared
with the  states of standard ghost structure.  With the physical states that
we have  described in section 2 and the appendix, there are many
non-vanishing  three-point functions.  We shall begin by presenting a
detailed computation of one example and shall then give the results for many
others.

    The easiest way to identify a possible non-vanishing three-point
function is first to ensure that the three states have momenta satisfying
the momentum-conservation law (3.5).  (As explained earlier, the only
non-trivial requirement comes from the conservation equation for the $\beta$
momentum.)  Since all states in a multiplet have the same momentum, it
suffices to look just at the prime state in each multiplet.  Having taken
care of momentum conservation, we must also ensure that the ghost structure of
the product of operators leads to the non-vanishing ghost correlation
function (3.1).  In particular, this means that the total ghost number
of the operators must be 8.  If the sum of ghost numbers for the three prime
states is greater than 8, then the three-point function will be zero.  If
the sum of ghost numbers equals 8, then this is a good candidate for a
non-vanishing three-point function.  If the sum is less than 8, then the
ghost numbers can be boosted by acting with $a_\varphi$ or $a_{X^\mu}^\dum$
to generate a higher ghost number member of a multiplet.  This again can
lead to a non-vanishing three-point function.

     Let us consider the following example in detail.  For the three prime
states we choose one to be a tachyon state with standard ghost structure,
with $\beta=-Q$; this corresponds to the operator $\V{G}{\deff}{\beta}{p}=
\V{3}{15/16}{-Q}{p_1}$ as
given in (A.3). The two remaining states that we choose have non-standard ghost
structure, and occur at level $\ell=1$.  One has $\beta=-\ft37Q$, and
corresponds to the operator $\V{2}{15/16}{-\ft37Q}{p_2}$ given in (A.7), and
the other has $\beta=-\ft47Q$ and corresponds to the operator
$\V{2}{1/2}{-\ft47Q}{p_3}$ given in (A.8).  We emphasise that even though
these are level 1 states, they are tachyonic from the point of view of the
matter, since the excitations are purely ghostly.  Clearly the $\beta$ momenta
satisfy the conservation condition (3.5).  However, the ghost numbers add up
to 7, and so we must boost this to 8 by acting on one of the prime states with
a linear combination of $a_\varphi$ and $a_{X^\mu}^\dum$.  We choose
to boost the ghost number of the second state; accordingly we take the
operator $\aV{G}{\deff}{\beta}{p} = \aV{3}{15/16}{-\ft37Q}{p_2}$ given in
(A.9).

     Not all the terms in these three states can combine to give the correct
ghost structure (3.4); only two of the three terms in $\aV{3}{15/16}{-\ft37Q}
{p_2}$, and one of the two terms in $\V{2}{1/2}{-\ft47Q}{p_3}$ contribute.
We find it convenient to compute the ghost and $\varphi$ part separately
from the effective matter part, since all the states that we consider in
this paper factorise in this way.  Thus we have for the ghost and $\varphi$
part
$$
\eqalign{
&\bra{0}\big(e^{i\sigma}e^{2i\rho}e^{-Q\varphi}\big) (z_1)\,
\big(11 e^{i\sigma}\del e^{2i\rho}e^{-\ft37Q\varphi}+
16 e^{i\sigma} e^{2i\rho} \del e^{-\ft37Q\varphi}\big) (z_2)\,
\big(e^{i\sigma}e^{i\rho}e^{-\ft47Q\varphi}\big) (z_3)\ket{0}\cr
&=- z_{12}z_{13}z_{23} \big[11 \del_2 (z_{12}^4 z_{13}^2 z_{23}^2)
z_{12}^{-21/8}
z_{13}^{-7/2} z_{23}^{-3/2}+
16 z_{12}^4 z_{13}^2 z_{23}^2 \del_2(z_{12}^{-21/8}
z_{13}^{-7/2} z_{23}^{-3/2})\big]\cr
&=-2\,z_{12}^{11/8} (z_{13}z_{23})^{1/2}\ ,\cr}\eqno(3.9)
$$
where $\del_i\equiv \del/\del z_i$.  The effective matter part is easier to
compute, since the operators are all tachyonic.  For this, we have
$$
\eqalign{
\big\langle e^{ip_1\cdot X(z_1)}\,e^{ip_2\cdot X(z_2)}\,e^{ip_3\cdot X(z_3)}
\big\rangle &=z_{12}^{p_1\cdot p_2}\,z_{13}^{p_1\cdot p_3}\,z_{23}^{p_2\cdot
p_3}\cr
&=z_{12}^{-11/8}(z_{13}z_{23})^{-1/2}\ ,\cr}\eqno(3.10)
$$
where in deriving the last line we have used the relation
$$
p_1\cdot p_2= \deff_3 -\deff_1 -\deff_2 =\ft12 -\ft{15}{16} -\ft{15}{16}
= - \ft{11}8\ ,\quad etc\ ,\eqno(3.11)
$$
which follows from the momentum-conservation law (3.5).  Putting the factors
(3.9) and (3.10) together, we finally obtain the three-point function
\medskip
\noindent{ $\bullet\
\underline{L^{\rm eff}_0=\{\ft{15}{16},\ \ft{15}{16},\ \ft12\}}$:}
$$
\bra{0}\V{3}{15/16}{-Q}{p_1}(z_1)\; \aV{3}{15/16}{-\ft37Q}{p_2}(z_2)\;
\V{2}{1/2}{-\ft47Q}{p_3}(z_3)\ket{0}= -2 \ .\eqno(3.12)
$$
That this three-point function is a constant is a consequence of the fact
that it is the correlator of three primary fields of dimension zero with
respect to the total energy-momentum tensor $T+T_{\rm gh}$.  The important
point is that the constant is non-zero, showing that there are indeed
interactions in the $W_3$ string.

    There are two three-point functions that are closely related to the one
that we have just calculated. These correspond to the cases where we boost
the ghost number of the first or the third operator instead of the second.
The computations are similar and we just give the results:
$$
\eqalign{
\bra{0}\aV{4}{15/16}{-Q}{p_1}(z_1)\; \V{2}{15/16}{-\ft37Q}{p_2}(z_2)\;
\V{2}{1/2}{-\ft47Q}{p_3}(z_3)\ket{0}&= -2 \ ,\cr
\bra{0}\V{3}{15/16}{-Q}{p_1}(z_1)\; \V{2}{15/16}{-\ft37Q}{p_2}(z_2)\;
\aV{3}{1/2}{-\ft47Q}{p_3}(z_3)\ket{0}&= -2 \ .\cr}\eqno(3.13)
$$
The fact that these two three-point functions also turn out to be non-zero
is a first sign of a pattern that we shall encounter in all the correlation
functions we compute, namely, that when it is necessary to boost the total
ghost number in a correlation function in order to reach 8, it does not seem
to matter which operator is boosted. (Recall that boosting an operator
means acting with a linear combination of $a_\varphi$ and $a_{X^\mu}^\dum$,
giving another member in the same multiplet, but with ghost number increased
by one.)

           Another interesting three-point function to compute is one with
three $L_0^{\rm eff}=1$ operators. The first two operators have standard ghost
structure with $\beta$-momenta $-\ft67 Q$ and $-\ft87 Q$. The third is a
level $\ell=3$ operator with non-standard ghost structure and $\beta=0$.
Their explicit expressions are given by (A.2), (A.1) and (A.12). The total
ghost number of these three operators is already 8, so no boosting is
needed. The result turns out to be
\medskip
\noindent{ $ \bullet\ \underline{L^{\rm eff}_0=\{1,\ 1,\ 1\}}$:}
$$
\bra{0}\V{3}{1}{-\ft67 Q}{p_1}(z_1)\; \V{3}{1}{-\ft87Q}{p_2}(z_2)\;
\V{2}{1}{0}{p_3}(z_3)\ket{0}= -2\ .\eqno(3.14)
$$
In the above calculation we have used
$$
\beta\, \del\gamma\, \gamma (z)=\,:\,e^{-i\rho} e^{2i\rho}\,:\,(z)=
{1\over {2\pi i}}
\oint {{d w}\over {w-z}} e^{-i\rho (w)} e^{2i\rho (z)}\ , \eqno(3.15)
$$
and similarly
$$
(\del \varphi)^2 (z)= {1\over{2\pi i}}\oint {{d w}\over {w-z}}\Big(
e^{-i\varphi (w)} \del e^{i\varphi (z)}-
\del  e^{-i\varphi (w)} e^{i\varphi (z)}  \Big)\ .\eqno(3.16)
$$
The relevant terms where these factors appear can then be evaluated as
four-point functions with a contour integral.

     From the physical operators given in the appendix, there are several
more  three-point functions that satisfy momentum conservation and have the
correct total ghost number. Classifying them by the $L^{\rm eff}_0$ values
of the operators,\footnote{$^*$}{\tenfoot The reason for classifying them
in this way is that, as we shall argue in section 4, physical operators with
the same $L_0^{\rm eff}$ value can be viewed as equivalent, even though they
may have different ghost structures and $\varphi$ dependence.} the results
are as follows:

\medskip
\noindent{ $\bullet \
\underline{L^{\rm eff}_0=\{\ft{15}{16},\ \ft{15}{16},\ 1\}}$:}
$$
\bra{0}\aV{3}{15/16}{-\ft37 Q}{p_1}(z_1)\; \V{2}{15/16}{-\ft37Q}{p_2}(z_2)\;
\V{3}{1}{-\ft87Q}{p_3}(z_3)\ket{0}=4 \ .\eqno(3.17)
$$
\bigskip
\noindent{ $\bullet\
\underline{L^{\rm eff}_0=\{\ft{15}{16},\ \ft{15}{16},\ 1\}}$:}
$$
\bra{0}\V{3}{15/16}{- Q}{p_1}(z_1)\; \V{3}{15/16}{-Q}{p_2}(z_2)\;
\V{2}{1}{0}{p_3}(z_3)\ket{0}=1 \ .\eqno(3.18)
$$
\bigskip
\noindent{ $\bullet\ \underline{L^{\rm eff}_0=\{\ft12,\ \ft12,\ 1\}}$:}
$$
\bra{0}\aV{3}{1/2}{-\ft47 Q}{p_1}(z_1)\; \V{2}{1/2}{-\ft27Q}{p_2}(z_2)\;
\V{3}{1}{-\ft87Q}{p_3}(z_3)\ket{0}= 2 \ .\eqno(3.19)
$$
\bigskip
\noindent{ $\bullet\ \underline{L^{\rm eff}_0=\{\ft12,\ \ft12,\ 1\}}$:}
$$
\bra{0}\aV{3}{1/2}{-\ft47 Q}{p_1}(z_1)\; \V{2}{1/2}{-\ft47Q}{p_2}(z_2)\;
\V{3}{1}{-\ft67Q}{p_3}(z_3)\ket{0}=4 \ .\eqno(3.20)
$$
\bigskip
\noindent{ $\bullet\ \underline{L^{\rm eff}_0=\{1,\ 1,\ \ft12\}}$:}
$$
\bra{0}\V{3}{1}{-\ft67 Q}{p_1}(z_1)\; \V{3}{1}{-\ft67Q}{p_2}(z_2)\;
\V{2}{1/2}{-\ft27Q}{p_3}(z_3)\ket{0}=0 \ .\eqno(3.21)
$$

   When the boosting of an operator was necessary in the above three-point
functions, we have only shown the results for one specific choice of which
operator to boost.  Similar expressions are obtained when a different
operator is  boosted.  No non-vanishing three-point functions where all three
physical operators have non-standard ghost structure can be constructed
with the examples given in the appendix.

     The vanishing of the three-point function (3.21) emerges only after a
computation.  The result is indicative of a relation with the fusion rules
of the Ising model, as indeed are the results of all the other three-point
functions given above.  We shall discuss this further in section 4.

\bigskip
\noindent {\it 3.4 The four-point function}
\bigskip

    By studying the poles of the four-point functions of the $W_3$ string, one
learns about the mass spectrum of theory.  There are several four-point
functions that we can calculate using the physical operators given in the
appendix.  We shall begin by describing the procedure for computing
four-point functions in the $W_3$ string, and illustrate it in detail with an
example. Then, we shall present the results for all the other four-point
functions.

      By making use of the $SL(2,C)$ invariance of the vacuum $\ket{0}$,
three of the worldsheet coordinates of the physical operators in a
four-point function may be fixed.  As usual, we choose to set $z_1=\infty$,
$z_2=1$ and $z_4=0$.  As in ordinary string theory the coordinate $z_3$
should then be integrated from 0 to 1, giving a scattering amplitude that is
independent of the positions of insertion of the physical operators.  The
physical operators have conformal dimension 0 with respect to the total
energy-momentum tensor $T+T_{\rm gh}$.  Therefore in order to preserve the
conformal covariance of the theory, the dimension of the physical operator
$V(z_3)$ inserted at $z_3$ must be increased by 1 so that we can integrate
over an operator of dimension 1, as we must for an invariant result.  This
may be achieved by making the replacement
$$
V(z_3)\rightarrow {1\over 2\pi i} \oint_{z_3} dw \, b(w)
V(z_3)\ ,\eqno(3.22)
$$
where the subscript on the contour integral indicates that it is to be
evaluated around a closed path enclosing $z_3$. The above procedure not only
preserves the projective structure (\ie\ $SL(2,C)$ covariance) but also
gives a result that is invariant under the BRST transformations generated by
(2.1).  This whole construction is parallel to the one used in ordinary
string theory.  It also admits an immediate generalisation to higher-point
functions.

     As in the case of three-point functions, a four-point function will be
zero unless the ghost numbers of the operators (after the $b$ contour
integral) add up to 8 and momentum
conservation is satisfied.  If these necessary conditions are satisfied then
it becomes a matter of computation to determine the result.  We shall now
carry this out for the following example. Let us take four physical
operators that all correspond to physical states with $L_0^{\rm eff}=\ft12$,
We can satisfy the momentum conservation by choosing three of them to be
level 1 states with non-standard ghost structure and $\beta=-\ft47 Q$, given
by (A.8), and
the other to be a level 2  state with non-standard ghost structure and
$\beta=-\ft27 Q$, given by (A.11). Since they already have the correct total
ghost number, to  insert the $b$ contour integral we need to boost one of the
operators to restore the total ghost number to 8. Without loss of
generality, we shall boost the operator at $z_4$. (As we shall
discuss later, it does not matter which operator we choose to boost.) Thus
the four-point function for these four operators takes the form:
$$
\eqalign{
&\int \! dz_3\oint_{z_3} \!{dw\over 2\pi i}\times \cr
&\bra{0} \V{2}{1/2}{-\ft47 Q}{p_1}(z_1)\; \V{2}{1/2}{-\ft47 Q}{p_2} (z_2) \,
b(w)\,\V{2}{1/2}{-\ft27 Q}{p_3}(z_3)\;  \aV{3}{1/2}{-\ft47
Q}{p_4}(z_4)\ket{0}\ ,\cr}\eqno(3.23)
$$
where the boosted operator $\aV{3}{1/2}{-\ft47 Q}{p_4}(z_4)$ is given in
(A.10).

     This four-point function may be factorised as a product of the
ghost plus $\varphi$ part, and a matter part.  For the ghost plus
$\varphi$ part,  we obtain
$$
\eqalign{
\bra{0}&\big(e^{i\sigma}e^{i\rho} e^{-\ft47 Q \varphi}\big)(z_1)
\big(e^{i\sigma}e^{i\rho} e^{-\ft47 Q \varphi}\big)(z_2)
\big(-\ft32 \del_3 e^{i\rho} e^{-\ft27 Q \varphi} -2 e^{i\rho} \del_3
e^{-\ft27Q\varphi}\big)(z_3)\cr
&\big( 10 e^{i\sigma}\del_4 e^{2i\rho} e^{-\ft47 Q\varphi} +12
 e^{i\sigma} e^{2i\rho} \del_4 e^{-\ft47 Q\varphi}\big)(z_4)\ket{0}\cr\cr
&=-15 {\cal A}_\sigma \del_3\del_4{\cal A}_\rho {\cal A}_\varphi
 -18{\cal A}_\sigma \del_3{\cal A}_\rho \del_4{\cal A}_\varphi
-20 {\cal A}_\sigma \del_4{\cal A}_\rho \del_3{\cal A}_\varphi
-24 {\cal A}_\sigma {\cal A}_\rho \del_3\del_4{\cal A}_\varphi\cr
&=2\, (z_{12}z_{14}z_{24})^{2/3} (z_{13}z_{23}z_{34})^{-1/3} x^{-2/3}
(1-x)^{-2/3} (1-x+x^2)\ .\cr}\eqno(3.24)
$$
Here ${\cal A}_\sigma$,  ${\cal A}_\rho$ and  ${\cal A}_\varphi$ denote the
contractions from the $\sigma$, $\rho$ and $\varphi$ fields, and are given
by
$$
\eqalign{
{\cal A}_\sigma &=z_{12}z_{14}z_{24}\ ,\cr
{\cal A}_\rho &= z_{12}z_{13}z_{23} z_{14}^2 z_{24}^2 z_{34}^2\ ,\cr
{\cal A}_\varphi &= z_{12}^{-2} z_{13}^{-1} z_{14}^{-2} z_{23}^{-1}
z_{24}^{-2} z_{34}^{-1}\ .\cr}\eqno(3.25)
$$
In the last line of (3.24) we have extracted the conformal prefactor, and
written the remainder in terms of the invariant cross-ratio
$$
x={z_{12} z_{34}\over z_{13} z_{24}}\ .\eqno(3.26)
$$

     In order to compute the effective matter part of the four-point
function, we introduce the Mandelstam variables $s$, $t$ and $u$:
$$
\eqalign{
s&\equiv -(p_1+p_2)^2 - 2 a \cdot (p_1+p_2)= -2p_1\cdot p_2 -2\Delta_1 -
2\Delta_2\ ,\cr
t&\equiv -(p_2+p_3)^2 - 2 a \cdot (p_2+p_3)=- 2p_2\cdot p_3 -2\Delta_2
-2\Delta_3\ ,\cr
u&\equiv -(p_1+p_3)^2 - 2 a \cdot (p_1+p_3)= -2 p_1\cdot p_3 -2\Delta_1
-2\Delta_3 \ .\cr}\eqno(3.27)
$$
Their sum is given by
$$
s+t+u=-2(\Delta_1+\Delta_2+\Delta_3+\Delta_4)\ .\eqno(3.28)
$$
In our present calculation, where $\Delta_i=\ft12$, we have $s+t+u=-4$.  The
matter part of the four-point function (3.23) gives
$$
\eqalign{
\big\langle e^{ip_1\cdot X(z_1)}\,e^{ip_2\cdot X(z_2)}\,e^{ip_3\cdot X(z_3)}
&\,e^{ip_4\cdot X(z_4)}
\big\rangle =z_{12}^{p_1\cdot p_2}\,z_{13}^{p_1\cdot p_3}\,
z_{14}^{p_1\cdot p_4}\,z_{23}^{p_2\cdot p_3}\,z_{24}^{p_2\cdot p_4}\,
z_{34}^{p_3\cdot p_4}\cr
&=(z_{12}z_{13}z_{14}z_{23}z_{24}z_{34})^{-1/3} x^{-s/2-2/3}(1-x)^{-t/2
-2/3}\ ,\cr}\eqno(3.29)
$$
where again we have extracted the appropriate conformal prefactor.
Combining (3.24) and (3.29), setting $z_1=\infty$, $z_2=1$ and $z_4=0$, and
integrating $x=z_3$ from 0 to 1, we finally obtain for the scattering
amplitude (3.23)
\medskip
\noindent{ $\bullet\ \underline{L^{\rm eff}_0=\{\ft{1}{2},\ \ft{1}{2},\
\ft{1}{2},\ \ft{1}{2}\}}$:}
$$
\eqalignno{\int\oint_{z_3}
\bra{0} &\V{2}{1/2}{-\ft47 Q}{p_1}(z_1)\; \V{2}{1/2}{-\ft47 Q}{p_2} (z_2)\,
b(w)\,\V{2}{1/2}{-\ft27
Q}{p_3}(z_3)\;  \aV{3}{1/2}{-\ft47 Q}{p_4}(z_4)\ket{0}\cr
&=2\, \int_0^1 \! dx \, x^{-s/2-2} (1-x)^{-t/2-2}(1-x+x^2)&(3.30)\cr
&=2\,{\Gamma(-s/2-1) \Gamma(-t/2)\over \Gamma(-s/2-t/2-1)}+ 2
{\Gamma(-s/2+1) \Gamma(-t/2-1)\over \Gamma(-s/2-t/2)} \ .
&(3.31)}
$$
Here, and in subsequent expressions, the integrals at the front of the first
line denote
precisely the integrations given in (3.23). We shall defer detailed
discussion of this result, and of the other four-point functions we shall
compute,  until section 4.

     There are several other four-point functions that satisfy momentum
conservation and have the correct ghost number, and we now list the results.
Just as for the three-point functions, we classify them by the $L_0^{\rm
eff}$  values of the physical operators.

\medskip
\noindent{ $\bullet\
\underline{L^{\rm eff}_0=\{\ft{15}{16},\ \ft{15}{16},\
\ft{1}{2},\ \ft{1}{2}\}}$:}
$$
\eqalignno{\int\oint_{z_3}
\bra{0} &\V{2}{15/16}{-\ft37 Q}{p_1}(z_1)\; \V{2}{15/16}{-\ft37 Q}{p_2} (z_2)
\,b(w)\,\aV{3}{1/2}{-\ft47
Q}{p_3}(z_3)\;  \V{2}{1/2}{-\ft47 Q}{p_4}(z_4)\ket{0}\cr
&=-2\int_0^1 \! dx \, x^{-s/2-2} (1-x)^{-t/2-31/16}(x-2)&(3.32)\cr
&=-2\,{\Gamma(-s/2) \Gamma(-t/2-15/16)\over \Gamma(-s/2-t/2-15/16)}+4\,
{\Gamma(-s/2-1) \Gamma(-t/2-15/16)\over \Gamma(-s/2-t/2-31/16)} \ .
&(3.33)}
$$
\bigskip

\medskip
\noindent{ $\bullet\
\underline{L^{\rm eff}_0=\{\ft{15}{16},\ \ft{15}{16},\
\ft{1}{2},\ \ft{1}{2}\}}$:}
$$
\eqalignno{\int\oint_{z_3}
\bra{0} &\V{3}{15/16}{- Q}{p_1}(z_1)\; \V{2}{15/16}{-\ft37 Q}{p_2} (z_2)
\,b(w)\,\V{2}{1/2}{-\ft27
Q}{p_3}(z_3)\;  \V{2}{1/2}{-\ft27 Q}{p_4}(z_4)\ket{0}\cr
&=0 \ .
&(3.34)}
$$
\bigskip

\medskip
\noindent{ $\bullet\
\underline{L^{\rm eff}_0=\{1,\ \ft{15}{16},\ \ft{15}{16},\
1 \}}$:}
$$
\eqalignno{\int\oint_{z_3}
\bra{0} &\V{3}{1}{-\ft87 Q}{p_1}(z_1)\; \V{2}{15/16}{-\ft37 Q}{p_2} (z_2)
\,b(w)\,\V{2}{15/16}{-\ft37
Q}{p_3}(z_3)\;  \V{2}{1}{0}{p_4}(z_4)\ket{0}\cr
&=  -2\, \int_0^1 \! dx \, x^{-s/2-31/16} (1-x)^{-t/2-2}&(3.35)\cr
&=-2\,{\Gamma(-s/2-15/16) \Gamma(-t/2-1)\over \Gamma(-s/2-t/2-31/16)}\ .
&(3.36)}
$$
\bigskip

\medskip
\noindent{ $\bullet\
\underline{L^{\rm eff}_0=\{\ft{15}{16},\ \ft{15}{16},\
1,\ \ft12\}}$:}
$$
\eqalignno{\int\oint_{z_3}
\bra{0} &\V{3}{15/16}{- Q}{p_1}(z_1)\; \V{2}{15/16}{-\ft37 Q}{p_2} (z_2)
\,b(w)\,\V{2}{1}{0}{p_3}(z_3)\;  \V{2}{1/2}{-\ft47Q}{p_4}(z_4)\ket{0}\cr
&=\int_0^1 \! dx \, x^{-s/2-3/2} (1-x)^{-t/2-31/16}&(3.37)\cr
&={\Gamma(-s/2-1/2) \Gamma(-t/2-15/16)\over \Gamma(-s/2-t/2-23/16)} \ .
&(3.38)}
$$
\bigskip

\medskip
\noindent{ $\bullet\
\underline{L^{\rm eff}_0=\{1,\ \ft{1}{2},\  1,\ \ft{1}{2}\}}$:}
$$
\eqalignno{\int\oint_{z_3}
\bra{0} &\V{3}{1}{-\ft67 Q}{p_1}(z_1)\; \V{2}{1/2}{-\ft47 Q}{p_2} (z_2)
\,b(w)\,\V{2}{1}{0}{p_3}(z_3)\;\V{2}{1/2}{-\ft47
Q}{p_4}(z_4)\ket{0} \cr
&=-2\int_0^1 \! dx \, x^{-s/2-3/2} (1-x)^{-t/2-3/2}&(3.39)\cr
&=-2\,{\Gamma(-s/2-1/2) \Gamma(-t/2-1/2)\over \Gamma(-s/2-t/2-1)}\ .
&(3.40)}
$$
\bigskip

\medskip
\noindent{ $\bullet\ \underline{L^{\rm eff}_0=\{1,\ \ft{1}{2},\
\ft{1}{2},\ 1\}}$:}
$$
\eqalignno{\int\oint_{z_3}
\bra{0} &\V{3}{1}{-\ft87 Q}{p_1}(z_1)\; \V{2}{1/2}{-\ft47 Q}{p_2} (z_2)
\,b(w)\,\V{2}{1/2}{-\ft27
Q}{p_3}(z_3)\;  \V{2}{1}{0}{p_4}(z_4)\ket{0}\cr
&=\int_0^1 \! dx \, x^{-s/2-3/2} (1-x)^{-t/2-2}&(3.41)\cr
&={\Gamma(-s/2-1/2) \Gamma(-t/2-1)\over \Gamma(-s/2-t/2-3/2)}\ .
&(3.42)}
$$
\bigskip

\medskip
\noindent{ $\bullet\ \underline{L^{\rm eff}_0=\{1,\ \ft12,\ \ft12,\ \ft12\}}$:}
$$
\eqalignno{\int\oint_{z_3}
\bra{0} &\V{3}{1}{-\ft67 Q}{p_1}(z_1)\; \V{2}{1/2}{-\ft47 Q}{p_2} (z_2)
\,b(w)\,\V{2}{1/2}{-\ft27
Q}{p_3}(z_3)\;  \V{2}{1/2}{-\ft27 Q}{p_4}(z_4)\ket{0}\cr
&=0\ .
&(3.43)}
$$
\bigskip

\medskip
\noindent{ $\bullet\
\underline{L^{\rm eff}_0=\{1,\ \ft12,\ \ft12,\ \ft12\}}$:}
$$
\eqalignno{\int\oint_{z_3}
\bra{0} &\V{3}{1}{-\ft87 Q}{p_1}(z_1)\; \V{2}{1/2}{-\ft27 Q}{p_2} (z_2)
\,b(w)\,\V{2}{1/2}{-\ft27
Q}{p_3}(z_3)\;  \V{2}{1/2}{-\ft27 Q}{p_4}(z_4)\ket{0}\cr
&=0\ .
&(3.44)}
$$
\bigskip

      In the four-point functions (3.30) and (3.32), where it is necessary
to boost the total ghost number, we have in each case made a specific
choice about which operator to boost. We have checked in several examples
that the result does not depend on which particular operator is chosen. We
expect this to be a general feature of all correlation functions.

     The four-point functions (3.34), (3.43) and (3.44) turn out to be zero
as a result of non-trivial computations.  The vanishing of (3.34) can be
understood from the underlying three-point functions, \ie\
that the intermediate state in the four-point function does not carry a
$\beta$ momentum for which there could be a non-vanishing three-point
function with the two external states for that particular channel.  This
suggests that the three-point function is indeed the basic interaction
vertex of the $W_3$ string.  The vanishing of the four-point functions
(3.43) and (3.44) cannot be explained in this way.  However, as we shall
discuss in section 4, it is indicative of a relation between the $W_3$
string and the Ising model.

\bigskip
\bigskip
\noindent
{\bf 4. Crossing, duality and the Ising model}
\bigskip

     Having obtained several non-vanishing three-point and four-point
functions in the previous section, we now turn to a discussion of the
significance of these results. We shall first discuss the crossing and
duality properties of these correlation functions.  Then we shall
investigate their relation with the fusion rules of the Ising model.

\bigskip
\noindent{\it 4.1 Crossing properties}
\bigskip

     One of the fundamental properties of the ordinary bosonic string is
that the four-point function is crossing symmetric [17]. For example, the
four-point function for tachyons in the ordinary string takes the form
$$
\int_0^1 dx\, x^{-s/2-2}\, (1-x)^{-t/2 -2}\ ,\eqno(4.1)
$$
and is invariant under the interchange of $s$ and $t$. It should be
stressed that this is a direct
consequence of the $SL(2,C)$ invariance of the vacuum $\ket{0}$. This
invariance in general dictates how correlation functions with different
orderings of the operators are related. In particular, for a case such as
(4.1), where the operators are identical, the four-point function is {\it
in}variant under $s\leftrightarrow t$, since the interchange of $s$ and $t$
corresponds precisely to the interchange of the second and fourth operators
or of the first and third operators.

     In the $W_3$ string there are analogous consequences of the $SL(2,C)$
invariance of the vacuum, again leading to relations among correlation
functions with different orderings of the operators. From these, we can see
that, as we mentioned in section 3, there are indications that the physical
operators we are discussing in this paper are characterised by their
$L_0^{\rm eff}$ values, regardless of their ghost structures and $\varphi$
dependence. For example, the four-point function (3.30) for four tachyons
with $L_0^{\rm eff}=\ft12$ is invariant under the interchange of $s$ and
$t$. So is the four-point function (3.39). Thus these four-point functions
exhibit the crossing symmetries that would be expected if the second and
fourth operators were identical and also if the first and the third
operators were identical. The fact that these pairs of operators in (3.30)
and (3.39) have identical $L_0^{\rm eff}$ values indeed provides evidence
for the above suggestion that all operators with an effective spacetime
interpretation and the same $L_0^{\rm eff}$ values should be regarded
as ``equivalent.''

    Since the above-mentioned pairs of operators  in the remaining
four-point functions of section 3 have different $L_0^{\rm eff}$ values,
they are not {\it in}variant under the ``crossing transformations.''
However, they do transform {\it co}variantly. For example, under the
crossing transformation $t\leftrightarrow u$, implemented by the
interchange of the third and fourth operators (and thus $x\to -x/(1-x)$),
the integrand in the four-point function (3.39) (including the measure $dx$)
transforms as
$$
dx \, x^{-s/2-3/2} (1-x)^{-t/2-3/2}\rightarrow
dx \, x^{-s/2-3/2} (1-x)^{-t/2-2}\ .\eqno(4.2)
$$
In fact, this result provides another illustration of the equivalence of
different physical operators with the same $L_0^{\rm eff}$ value, since
(4.2) leads precisely to (3.41).

\bigskip
\noindent{\it 4.2 Duality}
\bigskip

     Let us now turn to a discussion of duality. In the usual bosonic string,
duality is the statement that the four-point amplitude can be expanded as an
infinite sum over either $s$-channel or $t$-channel poles [17]. The form of
the poles in the sum is identical in each case, since it is always the same
set of intermediate string states that are exchanged in either channel.

    The notion of duality exists for the $W_3$ string, but it is slightly
more subtle. In this case, four-point amplitudes can again be expanded as
infinite sums over either $s$-channel or $t$-channel poles, as immediately
follows from their expressions in terms of $\Gamma$ functions. However, as
we have seen, the $W_3$ string contains different sectors corresponding to
different $L_0^{\rm eff}$ values, and the set of intermediate states in one
channel does not necessarily belong to the same sector as the set of
intermediate states in another channel. Thus the form of the poles may
be different in different channels. In particular, the masses of the
exchanged particles in different channels may not be the same.

    The four-point function (3.32) provides a nice illustration of this
phenomenon. As can be seen from (3.33), if we expand in the $s$-channel, the
poles correspond to the exchange of states from the  $L_0^{\rm eff}=1$
sector, with $({\rm mass})^2=2n-2$, whilst if we expand in the $t$-channel,
the poles correspond to the exchange of states from the  $L_0^{\rm
eff}=\ft{15}{16}$ sector, with $({\rm mass})^2=2n-\ft{15}8$. Similarly, in
the four-point function (3.37), the exchanged states in the $s$-channel have
$L_0^{\rm eff}=\ft12$, with $({\rm mass})^2=2n-1$, whereas in the
$t$-channel the exchanged states have $L_0^{\rm eff}=\ft{15}{16}$.  Our
expressions for four-point functions  in subsection 3.4 make manifest the
structure of poles in the $s$-channel and  $t$-channel.  Of course one can
also look at the $u$-channel, where the  exchanged states may again belong
to a different sector.  For example, the  four-point function (3.39) is
$s\leftrightarrow t$ symmetric, with $L_0^{\rm  eff}=\ft12$ states exchanged
in both these channels; if expanded instead in  the $u$-channel, the
exchanged states have $L_0^{\rm eff}=1$.  The other four-point functions in
section 3 provide further examples.

\bigskip
\noindent{\it 4.3  The relation with the Ising model}
\bigskip

     Indications of a relation with the Ising model have been apparent since
the earliest work on the $W_3$ string [1,2,5].  Until now, the evidence was
essentially numerological, consisting of a twofold observation. Firstly, the
central charge of the effective energy-momentum tensor $T^{\rm eff}$ given
in (2.5) can be written as $\ft{51}{2}=26-\ft12$, where 26 is the critical
central charge of the usual bosonic string and $\ft12$ is the central charge
of the Ising model. Secondly, the set of $L_0^{\rm eff}$ values, namely
$\{1,\ \ft{15}{16},\ \ft12\}$, can be written as $1-\Delta_{\rm min}$, where
1 is the intercept of the usual bosonic string and $\Delta_{\rm min}$ takes
the values of the dimensions of the primary fields of the Ising model, namely
$\{0,\ \ft1{16}, \ \ft12\}$.

    Inspired by this numerological connection, it was recently proposed in
[18] that one might be able to compute the scattering amplitudes for those
physical states of the $W_3$ string that admit an effective spacetime
interpretation by tensoring, by hand, the effective spacetime parts of the
physical states with appropriate primary fields of the Ising model, and
then calculating the scattering amplitudes for the tensor-product states.
This  procedure implicitly assumes that the physical operators of the $W_3$
string  are equivalent to direct products of effective Virasoro operators
with Ising  fields.   It is {\it a priori} far from clear that this direct
product structure captures the essence of the $W_3$ symmetry.  Later in this
subsection we shall examine this assumption in more detail in the light of
our calculation of the $W_3$-string scattering amplitudes, and show that
indeed there is more to the $W_3$ symmetry than can be described by a
direct-product structure.

     It follows from our results in section 3 that the connection between
the $W_3$ string and the Ising model is more than numerological. In fact,
the pattern of vanishing and non-vanishing three-point functions computed
in subsection 3.3 reproduces the fusion rules of the Ising model. To see
this, we associate, as suggested by the numerological observation, the
$L_0^{\rm eff}=1$ sector with the identity operator {\bf 1} of the Ising
model; the
$L_0^{\rm eff}=\ft{15}{16}$ sector with the spin operator $\sigma$; and the
$L_0^{\rm eff}=\ft12$ sector with the energy operator $\varepsilon$. It is
now immediately clear that the structure of the three-point functions
presented in subsection 3.3 exactly agrees with the fusion rules [19] of the
Ising model, {\it viz.}
$$
\eqalign{
{\bf 1}\times {\bf 1}&={\bf 1}\ ,\qquad \sigma\times\sigma={\bf
1}+\varepsilon\ ,\cr
{\bf 1}\times \sigma&=\sigma\ ,\qquad \sigma\times\varepsilon=\sigma\ ,\cr
{\bf 1}\times \varepsilon&=\varepsilon\ ,\qquad
 \varepsilon\times \varepsilon= {\bf 1}\ .\cr}\eqno(4.3)
$$

     As we already mentioned, the three-point function describes the basic
interaction vertex in the $W_3$ string. Thus one may expect, and indeed it
is the case, that all the four-point functions that we have computed should
be consistent with the above fusion rules and the corresponding four-point
functions of the Ising model, in the sense that an amplitude that is
forbidden by the fusion rules (4.3) is indeed zero in the $W_3$ string. In
particular, this explains the vanishing of  (3.43) and (3.44). However, it
appears that this correspondence does not go  both ways. From the fusion
rules of the Ising model, one might expect that  there should exist
four-point functions for four $L_0^{\rm eff}=1$ operators  and also for four
$L_0^{\rm eff}=\ft{15}{16}$ operators.  With the physical  operators given
in this paper, we cannot construct either of these; in  neither case it is
possible to have the correct total ghost structure together with momentum
conservation in the $\varphi$ direction.  For the  $L_0^{\rm eff}=1$ case,
this may simply be because we have not explored  higher-level states in the
physical spectrum sufficiently. However, for the   $L_0^{\rm
eff}=\ft{15}{16}$ case on the other hand, it seems impossible, from the
general pattern of physical states that is known so far, to construct such
a  four-point function satisfying $\beta$-momentum conservation. To see
this, suppose that, as it is the case for all known physical states, the
$\beta$ momentum is of the form $\beta= \ft{k}7 Q$, with $k$ an
integer.\footnote{$^*$}{\tenfoot In fact the quantisation of $\beta$ in
units of $\ft17 Q$ follows from a functional integral approach if $\varphi$
is taken as a time-like coordinate, since under this circumstance $\varphi$
is automatically periodic [4,5].} It follows that a physical state with
$L_0^{\rm eff}=\ft{15}{16}$ would have $k=-7\pm 4 \sqrt n$, where $n$ is an
integer related to the level number. To satisfy momentum conservation in the
four-point function, we would need four such (integer) $k$'s satisfying
$k_1+k_2+k_3+k_4=-14 $, which is manifestly impossible. (In fact even if one
relaxes the supposition that the $k$'s are integers, it would still be the
case that momentum conservation could not be satisfied for such a four-point
function. This follows from the relatively-easily proven fact that, for
integer $n_i$, the sum $\pm\sqrt{n_1}
\pm\sqrt{n_2}\pm\sqrt{n_3}\pm\sqrt{n_4}$ cannot equal $\ft72$.)

     Let us now compare this result with the computation in [18] of the
four-point function for four $\leff_0=\ft{15}{16}$ effective Virasoro
operators tensored with four spin-$\ft1{16}$ fields of the Ising model.
Interestingly, this computation gave a non-zero result.  In view of our
finding above, namely that the four-point function of four
$\leff_0=\ft{15}{16}$ physical operators in the $W_3$ string is zero, it
seems that essential aspects of the $W_3$ symmetry are not captured by the
method of [18].  It appears from our results that a $W_3$-symmetry
selection rule forbids the existence of such a non-vanishing four-point
function in the $W_3$ string.

    We have checked that if one uses the method in [18] to calculate the
four-point function for four $L_0^{\rm eff}=\ft12$ effective Virasoro
operators tensored with four spin-$\ft12$ Ising fields, the result is
the same as the result (3.30) for the scattering of four $\leff_0=\ft12$
physical states of the $W_3$ string. It is not clear to us to what extent in
general the method presented in [18] should agree with the $W_3$-string
scattering amplitudes, which are described in this paper.

\np
\noindent{\bf 5. Open problems and conclusions}
\bigskip

     In this paper we have presented a procedure for computing
gauge-invariant scattering  amplitudes in the $W_3$ string. Although we have
concentrated on the open  string, the procedure is equally applicable to the
closed string. The  essential point that enables us to build scattering
amplitudes is the existence of physical states with non-standard ghost
structure; their inclusion is vital, since there seems to be no way  to
obtain non-vanishing scattering amplitudes amongst physical states with
only standard ghost structure.

    All the physical states we have considered in this paper have the
property of factorising into a product of the form
$$
\ket{\rm effective\  spacetime}\,\otimes\, \ket{{\rm ghost}+\varphi}\ .
\eqno(5.1)
$$
We have observed in several examples that the $\leff_0$ value characterises
these physical states, in the sense that states with the same $\leff_0$ but
different ghost structure and $\varphi$ dependence behave equivalently in
correlation functions.

    For simplicity, we have restricted our attention to physical states of
the form (5.1) where $\ket{\rm effective\  spacetime}$ is a tachyonic state.
The only property of $\ket{\rm effective \ spacetime}$ that is relevant when
imposing the physical state conditions on a state of the form (5.1) is that
it should be a highest-weight state under $T^{\rm eff}$ with weight
$\leff_0$. This means that we can replace the effective tachyonic state by
an arbitrary excited effective physical state with the same intercept
$\leff_0$. One can then straightforwardly write down scattering amplitudes
for these new excited physical states.

     There are, however, many states in the physical spectrum of the
$W_3$ string that are not of the form (5.1).  Specifically, they have
prefactors that are the sum of terms that involve $X^\mu$ excitations and
terms that do not, and thus they do not factorise into the form (5.1).  The
procedure that we have developed in this paper for calculating scattering
amplitudes is equally applicable for such states.  The effective spacetime
interpretation of these physical states, and of the corresponding scattering
amplitudes, is not yet clear.  In addition, it is not clear how, if at all,
these non-factorisable states are related to the Ising model.

     An important issue that has not been addressed in this paper is the
question of unitarity.  For physical states with standard ghost structure,
unitarity was proven in [5], by exploiting the fact that they all admit an
effective spacetime interpretation with effective intercept values
$\leff_0=\ft{15}{16}$ or $\leff_0=1$.  One can easily show that  the
unitarity bounds for an effective Virasoro string with central charge
$c=\ft{51}2$ imply that the effective intercept must satisfy either
$\ft{15}{16} \le \leff_0\le 1$ or $0\le \leff_0\le \ft12$ [5].  Thus the
standard-type physical states of the $W_3$ string precisely saturate the
lower and upper limits of the first of these unitarity bounds.  In fact {\it
all} the known physical states of the $W_3$ string that admit an effective
spacetime interpretation saturate one or other of the unitarity bounds,
since they have $\leff_0=1$, $\ft{15}{16}$ or $\ft12$. Further work on the
question of unitarity for the $W_3$ string is in progress.

     We restricted attention in this paper to physical operators at ghost
number $G=3$ (for operators with standard ghost structure), or $G=2$ (for
operators with non-standard ghost structure), and their boosted partners at
$G=4$ or $G=3$ respectively.  It is easy to see that when $N$ is greater
than 5, $N$-point functions of such operators can never have the correct
ghost structure, and thus they all vanish.  However, as we mentioned in
section 2, there are also physical operators with ghost number $G=0$; the
first non-trivial example (\ie\ with {\it continuous} on-shell spacetime
momentum $p_\mu$) occurs at level $\ell=8$, with $\beta=\ft47Q$ [16].
Further examples will arise at higher levels, and can be generated by the
action of the ground-ring generators of the $W_3$ string [10,16].  With such
states the $N\le5$ limit discussed above can clearly be overcome.  It is
interesting to note also that these $G=0$ operators have positive $\beta$
momentum, and can thus counterbalance the negative contributions from the
$G=2$ and $G=3$ operators discussed in this paper.

     The procedure that we have developed in this paper seems to provide a
consistent picture of $W_3$-string scattering.  Although we can calculate
any desired scattering amplitude, the underlying structure remains obscure.
This is not surprising, since elucidating the organising principle  would
require the understanding of $W$ geometry.  It may be, however, that  our
results provide us with a glimpse of $W$ geometry itself.  By analogy with
the super-extension of the bosonic string, where one introduces superspace
and integrates over it, one should expect that in a $W$ extension of the
bosonic string one should introduce a ``$W$ space'' and ``integrate'' over
it.  In view of our results it is not inconceivable that this integration
over $W$ space turns states with standard ghost structure into  states with
non-standard ghost structure and {\it vice versa}.  This suggests that $W$
geometry cannot be completely understood without including ghosts.

     In a usual gauge theory the r\^ole of the ghost fields is to remove
the unphysical degrees of freedom of the gauge fields.  As such, they appear
only as virtual particles, and never as external states in physical
amplitudes.  In the $W_3$ string, on the other hand, their r\^ole seems to
be strikingly different: ghost excitations must necessarily appear in the
external states of physical amplitudes.  Whereas ghost fields in external
states would spoil unitarity in a usual gauge theory, such as Yang-Mills, it
seems plausible that they are {\it needed} for unitarity of the $W_3$
string.  Indeed the $W_3$ string differs in an essential way from a usual
gauge theory in that the gauge algebra of the matter fields in the quantum
theory, namely $W_3$, is different from the gauge algebra of the original
classical theory, which is a contraction of $W_3$.  This non-trivial
renormalisation of the gauge algebra does not happen in a usual gauge
theory.  It may well be, therefore, that it is inappropriate to try to
understand $W_3$-string theory from a classical point  of view.

\np

\centerline{\bf ACKNOWLEDGMENTS}
\bigskip

     We are grateful to John Dixon, Mike Duff, HoSeong La, Kelly Stelle and
Peter West for  discussions.  Stany Schrans thanks the Center for Theoretical
Physics at  Texas A\&M University for hospitality and support.  We have made
extensive  use of the Mathematica package OPEdefs [20] written by Kris
Thielemans.

\vskip1truein

\noindent
\centerline{\bf APPENDIX}
\bigskip
In this appendix we present the explicit forms of the various physical
operators that we use in this paper.  In particular, we give the operators
corresponding to all the prime states at level 0 (\ie\ with standard ghost
structure), level 1 and level 2, and one example at level 3 (these all have
non-standard ghost structure). We recall that we denote such an operator by
$\V{G}{\Delta}{\beta}{p}$, where $G$ is its ghost number, $\Delta$ is its
$\leff_0$ value, and $\beta$ and $p$ are the momenta in the $\varphi$ and
effective spacetime ($X^\mu$) directions.  For level 0 and level 1, we also
give for each prime state a member of the multiplet whose ghost number has
been boosted by 1.  Such operators are denoted by
$\aV{G}{\Delta}{\beta}{p}$.  In each case we choose a special linear
combination of the ghost boosters $a_\varphi$ and $a_{X^\mu}^\dum$ with
which we act on the prime state.  This combination is chosen so as to
introduce no excitations in the $X^\mu$ directions, in order to preserve the
factorisability (5.1) and consequently to permit the same effective
spacetime interpretation for the boosted physical states that the prime
states enjoy.  Expressions for $a_\varphi$ and $a_{X^\mu}^\dum$ are
presented in [10] for the two-scalar case, and can be generalised immediately
to the multi-scalar case.

\noindent$\bullet\ ${\it Standard ghost structure: level 0}
$$\eqalignno{
\V{3}{1}{-\ft87 Q}{p} &=c\, \del \gamma\, \gamma\, e^{-\ft87 Q \varphi}
e^{i p\cdot X},&(A.1)\cr
\V{3}{1}{-\ft67 Q}{p} &=c\, \del \gamma\,\gamma \, e^{-\ft67 Q \varphi}
e^{i p\cdot X},
&(A.2)\cr
\V{3}{15/16}{-Q}{p} &=c\, \del\gamma\,\gamma \, e^{-Q \varphi}
e^{i p\cdot X}.&(A.3) \cr}
$$
$\bullet\ ${\it Boosted operators}
$$\eqalignno{
\aV{4}{1}{-\ft87 Q}{p} &=c\,\del^2 \gamma\,\del \gamma\, \gamma\,
e^{-\ft87 Q \varphi} e^{i p\cdot X},&(A.4)\cr
\aV{4}{1}{-\ft67 Q}{p} &=c\, \del^2\gamma\,\del \gamma\,\gamma\,
e^{-\ft67 Q \varphi} e^{i p\cdot X},
&(A.5)\cr
\aV{4}{15/16}{-Q}{p} &= c\,\del^2\gamma\,\del\gamma\,\gamma\,
e^{-Q \varphi} e^{i p\cdot X}.&(A.6) \cr}
$$
$\bullet\ ${\it Non-standard ghost structure: level 1}
$$\eqalignno{
\V{2}{15/16}{-\ft37 Q}{p} &=\Big(c\,\gamma + {i \over 3
\sqrt{58}}\del\gamma\,  \gamma\Big)\, e^{-\ft37 Q \varphi} e^{ip\cdot X}
,&(A.7)\cr \V{2}{1/2}{-\ft47 Q}{p} &=\Big(c\,\gamma - {i \over 3
\sqrt{58}}\del\gamma\,  \gamma\Big)\, e^{-\ft47 Q \varphi} e^{ip\cdot X}
.&(A.8)\cr} $$
$\bullet\ ${\it Boosted operators}
$$\eqalignno{
\aV{3}{15/16}{-\ft37 Q}{p} &=\Big(11c\,\del^2\gamma\,\gamma-12 \sqrt{2}
\del\varphi\, c\,
\, \del\gamma\,\gamma - {13 i\over \sqrt{58}} \del^2\gamma\,\del\gamma\,
\gamma  \Big)\, e^{-\ft37 Q \varphi} e^{ip\cdot X} ,&(A.9)\cr
\aV{3}{1/2}{-\ft47 Q}{p} &= \Big(10c\,\del^2\gamma\,\gamma-12 \sqrt{2}
\del\varphi\,c\,
\, \del\gamma\,\gamma - {38 i\over 3\sqrt{58}} \del^2\gamma\,\del\gamma\,
\gamma  \Big)  \, e^{-\ft47 Q \varphi} e^{ip\cdot X}\ .
&(A.10)\cr}
$$
$\bullet\ ${\it Non-standard ghost structure: level 2}
$$\eqalign{
\V{2}{1/2}{-\ft27 Q}{p} = &{-i \over \sqrt{29}}
\Big(\del\varphi\,\del\gamma\,\gamma
+ \sqrt{58}\, i\,\del\varphi \,c\,\gamma
-\ft32 \sqrt{29}\, i\, c\,\del\gamma   \cr
&-\ft23 \sqrt{2}
\del^2\gamma\,\gamma - \ft13 \sqrt{2}\, b\,c\,\del\gamma\,\gamma\Big)
e^{-\ft27 Q \varphi} e^{i p\cdot X}.\cr}\eqno(A.11)
$$
$\bullet\ ${\it Non-standard ghost structure: level 3}
$$
\eqalign{
\V{2}{1}{0}{p}=
&\Big(\ft{680}{261}b\,\del^2\gamma\,\del\gamma\,\gamma - 36
c\,\beta \,\del
\gamma\,\gamma - 19 c\,\del^2\gamma - \ft{16}{29}\sqrt{29}\, i\,
\del\varphi\,b\,c\,\del\gamma\,\gamma +42 \sqrt{2}\, \del
\varphi\,
c\,\del\gamma \cr
&-24 (\del\varphi)^2 c\,\gamma + \ft{28}{29}\sqrt{58} \,i\,
(\del\varphi)^2 \del\gamma\,\gamma - \ft{172}{87}\sqrt{29}\, i
\,\del\varphi  \,\del^2\gamma\,\gamma -\ft{14}{29}\sqrt{58}\, i\, \del
b\,c\,\del\gamma\, \gamma \cr
& + \ft{48}{29}\sqrt{29} \,i\, \del^2\varphi\, \del\gamma\,\gamma +
\ft{301}{174}\sqrt{58}\,i\, \del^2\gamma\,\del\gamma\Big)e^{ip\cdot X}.\cr}
\eqno(A.12)
$$

\np

\singlespace
\centerline{\bf REFERENCES}
\frenchspacing
\bigskip

\item{[1]}A. Bilal and J.-L. Gervais, {\sl Nucl. Phys.} {\bf B314} (1989)
646.

\item{[2]}S.R. Das, A. Dhar and S.K. Rama, {\sl Mod. Phys. Lett.}
{\bf A6} (1991) 3055; {\sl Int. J. Mod. Phys.} {\bf A7} (1992) 2295.

\item{[3]}C.N. Pope, L.J. Romans and K.S. Stelle, {\sl Phys.
Lett.} {\bf B268} (1991) 167; {\sl Phys. Lett.} {\bf B269} (1991) 287.

\item{[4]}C.N.\ Pope, L.J.\ Romans, E.\ Sezgin and K.S.\ Stelle,
{\sl Phys. Lett.} {\bf B274} (1992) 298.

\item{[5]}H. Lu, C.N. Pope, S. Schrans and K.W. Xu,  {\sl Nucl.
Phys.} {\bf B385} (1992) 99.

\item{[6]}H. Lu, C.N. Pope, S. Schrans and X.J. Wang,  {\sl Nucl.
Phys.} {\bf B379} (1992) 47.

\item{[7]}J. Thierry-Mieg, {\sl Phys. Lett.} {\bf B197} (1987) 368.

\item{[8]}H. Lu, B.E.W. Nilsson, C.N. Pope, K.S. Stelle and P.C. West,
``The low-level spectrum of the $W_3$ string,'' preprint CTP TAMU-64/92.

\item{[9]}S.K. Rama, {\sl Mod.\ Phys.\ Lett.}\ {\bf A6} (1991) 3531.

\item{[10]}C.N. Pope, E. Sezgin, K.S. Stelle and X.J. Wang, ``Discrete
States in the $W_3$ String,'' preprint, CTP TAMU-64/92,
Imperial/TP/91-92/40, hep-th/9209111, to appear in {\sl Phys. Lett}.
{\bf B}.

\item{[11]}B.H. Lian and G.J. Zuckermann, {\sl Phys. Lett.}\ {\bf B254} (1991)
 541.

\item{[12]}E. Witten, {\sl Nucl. Phys.} {\bf B373} (1992) 187;\nl
E. Witten and B. Zwiebach, {\sl Nucl. Phys.} {\bf B377} (1992) 55.

\item{[13]}L.J.  Romans, {\sl Nucl.  Phys.} {\bf B352} (1991) 829.

\item{[14]}P.C. West, ``On the spectrum, no ghost theorem and modular
invariance of $W_3$ strings,'' preprint KCL-TH-92-7.

\item{[15]}H. Lu, C.N. Pope and P.C. West, work in progress.

\item{[16]}C.N. Pope and X.J. Wang, preprint in preparation.

\item{[17]}M.B. Green, J.H. Schwarz and E. Witten, ``Superstring theory,''
Cambridge University Press, 1987.

\item{[18]}M.D. Freeman and P.C. West, ``$W_3$ String Scattering,''
preprint, KCL-TH-92-4.

\item{[19]}A.A. Belavin, A.M. Polyakov and A.B. Zamolodchikov, {\sl Nucl.
Phys.} {\bf B241} (1984) 333.

\item{[20]}K. Thielemans, {\sl Int. J, Mod. Phys.} {\bf C2} (1991) 787.

\end